\documentclass[%
reprint,
amsmath,
amssymb,
aps,
pra,
longbibliography
]{revtex4-2}

\usepackage{graphicx}
\usepackage{dcolumn}
\usepackage{bm}

\usepackage{mathtools}

\usepackage{color}

\newcommand{\nc}{\newcommand}
\nc{\bra}{\langle}
\nc{\ket}{\rangle}
\nc{\vac}{|0\ket}
\nc{\da}{^{\dagger}}
\nc{\HASEP}{\mathcal{H}}

\nc{\so}{\hat{S}}
\nc{\nm}{\hat{n}}
\nc{\pt}{\Tilde{P}}

\nc{\im}{\text{i}}
\nc{\Imath}{\mathcal{I}}
\nc{\blue}{\textcolor{blue}}
\nc{\red}{\textcolor{red}}

\nc{\la}{\lambda}
\nc{\al}{\alpha}
\nc{\ze}{\zeta}

\begin{document}

\title{Asymmetric simple exclusion process with tree-like network branches}

\author{Yuki Ishiguro$^{1,2,3}$}
\author{Yasunobu Ando$^3$}
 \affiliation{
  $^1$Faculty of Engineering, Tokyo Polytechnic University, 5-45-1 Iiyama-minami, Atsugi, Kanagawa 243-0297, Japan\\
 $^2$The Institute for Solid State Physics, The University of Tokyo, 5-1-5 Kashiwanoha, Kashiwa, Chiba 277-8581, Japan\\
$^3$Laboratory for Chemistry and Life Science, Institute of Science Tokyo, 4259 Nagatsuta-cho, Midori-ku, Yokohama 226-8503, Japan
 }

\begin{abstract}
The asymmetric simple exclusion process (ASEP) is a fundamental stochastic model describing asymmetric many-particle diffusion with hard-core interactions on a one-dimensional lattice, and has been widely applied in the study of nonequilibrium transport phenomena.
Motivated by the modeling of proton transport along oxygen networks in proton-conducting solid oxides, we extend the ASEP to a model defined on a one-dimensional backbone lattice with tree-like network branches. 
We derive the exact stationary distribution of this network ASEP and investigate its transport properties. 
By considering two representative network geometries for which physical quantities can be expressed in terms of certain hypergeometric series, we elucidate how the network geometry influences transport properties.
\end{abstract}
\maketitle


\section{Introduction}
Understanding and designing transport in many-particle systems is a fundamental problem across a broad range of applications.
While interparticle interactions strongly influence transport properties, the analysis of many-particle systems is often challenging. 
The asymmetric simple exclusion process (ASEP) is a paradigmatic model for describing nonequilibrium transport in many-particle systems with hard-core interactions \cite{Derrida_1993,derrida1998exactly,blythe2007nonequilibrium,Crampe_2014,essler1996representations,golinelli2006asymmetric,gwa1992bethe,kim1995bethe,Golinelli_2004,Golinelli_2005,PhysRevE.85.042105,Motegi_2012,prolhac2013spectrum,prolhac2014spectrum,prolhac2016extrapolation,prolhac2017perturbative,ishiguro2023,de2005bethe,deGier_2006,deGier_2008,deGier_2011,Wen_2015,Crampe_2015,Sandow_1994,relation,schadschneider2000statistical,schadschneider2010stochastic,macdonald1968kinetics,klumpp2003traffic}. 
Although the ASEP is a simple model in which many particles with hard-core interactions perform asymmetric random walks on a one-dimensional (1D) lattice, it captures the essence of transport phenomena under these interactions and has been applied to a wide range of nonequilibrium phenomena, such as traffic flow \cite{schadschneider2000statistical,schadschneider2010stochastic} and biological transport \cite{macdonald1968kinetics,klumpp2003traffic}. 
A particularly remarkable feature of the ASEP is its exact solvability, which allows for the exact evaluation of physical quantities using techniques from mathematical physics, such as the matrix product ansatz \cite{Derrida_1993,derrida1998exactly,blythe2007nonequilibrium,Crampe_2014,essler1996representations} and the Bethe ansatz \cite{golinelli2006asymmetric,gwa1992bethe,kim1995bethe,Golinelli_2004,Golinelli_2005,PhysRevE.85.042105,Motegi_2012,prolhac2013spectrum,prolhac2014spectrum,prolhac2016extrapolation,prolhac2017perturbative,ishiguro2023,de2005bethe,deGier_2006,deGier_2008,deGier_2011,Wen_2015,Crampe_2015}. 

While the 1D ASEP has been studied extensively, extending the ASEP to more complex settings is also an important challenge to address more diverse and realistic phenomena. In particular, by extending the lattice on which the ASEP is defined to situations more complex than a 1D lattice, such as higher-dimensional lattices \cite{alexander1992shock,Cai_2008,SINGH20093113,SHI20122640,DING20181700,Curatolo_2016,PhysRevResearch.6.033030,Ishiguro_2025,PhysRevE.84.061141,wang2017dynamics,wang2018analytical} and networks \cite{PhysRevE.69.066128,Pronina_2005,PhysRevE.77.051108,PhysRevE.80.041128,Basu_2010,PhysRevLett.107.068702,Neri_2013,Mottishaw_2013,PhysRevLett.134.027102,Ezaki_2012,Sarkar_2025}, one can describe a broader class of phenomena, including multilane traffic flow. In such complex settings, analyses are often carried out using mean-field approximations. However, several exactly solvable cases are also known, even in higher-dimensional \cite{PhysRevResearch.6.033030,Ishiguro_2025,PhysRevE.84.061141,wang2017dynamics,wang2018analytical} and network extensions \cite{Ezaki_2012,Sarkar_2025}.

In this work, we extend the ASEP with the aim of applying it to the study of ionic transport in solids, particularly to the optimal design of proton transport along oxygen networks in proton-conducting solid oxides.
Solid electrolytes are ion-conducting solids that have attracted attention as materials with promising applications in a wide range of technologies, including all-solid-state batteries and fuel cells \cite{bachman2016inorganic,famprikis2019fundamentals,zhao2020designing,hossain2017review,hussain2020review}. In recent years, theoretical studies using exclusion processes have been proposed to elucidate the collective aspects of ionic transport in such materials \cite{PhysRevResearch.6.L032032,PhysRevResearch.6.043210,PhysRevResearch.7.023068}.
This work focuses on transport phenomena in proton-conducting solid oxides, which constitute a class of solid electrolytes \cite{10.1063/1.5135319,proton2}. In solid oxides, a network composed of oxygen atoms is formed, and protons are considered to migrate primarily along this oxygen network. Since each oxygen typically binds at most one proton, the transport of protons can be modeled as an exclusion process by regarding oxygens as lattice sites and protons as particles.
The geometry of the oxygen network depends on the material. Identifying materials that realize ideal properties from a vast number of candidates remains an important challenge. If the ionic transport properties corresponding to the network geometry can be understood, such knowledge can be applied to the screening of candidate materials.

In this paper, we introduce the ASEP on a 1D backbone lattice with tree-like network branches as a model of proton transport along oxygen networks (Fig. \ref{fig:asep_fig}(a)), and investigate how network geometry affects transport properties. 
By applying the method proposed in \cite{PhysRevResearch.6.033030}, we derive the exact steationary distribution of this network ASEP.
For two characteristic network geometries introduced later (Fig. \ref{fig:substructures}), physical quantities can be expressed in terms of certain hypergeometric series. Using these examples, we perform exact analyses and clarify the effect of the network geometries on transport properties.

This paper is organized as follows. 
In Sec. \ref{sec:model}, we introduce the ASEP with tree-like network branches. We define the model and formulate the master equation describing the time evolution of the system.
In Sec. \ref{sec:steady}, we derive the exact steationary distribution of the model.
As a preparation for deriving the exact stationary distribution of the ASEP with tree substructures (Fig. \ref{fig:asep_fig}(c)), we first review the stationary distribution of the ASEP under periodic boundary conditions (Fig. \ref{fig:asep_fig}(a)) and derive the stationary distribution of the ASEP on tree structures (Fig. \ref{fig:asep_fig}(b)). Subsequently, we obtain the stationary distribution of the ASEP with tree substructures (Fig. \ref{fig:asep_fig}(c)) by applying the strategy proposed in  \cite{PhysRevResearch.6.033030}.
In Sec. \ref{sec:transport}, we investigate transport properties of the model. 
In this section, we focus on the two representative network geometries (Fig. \ref{fig:substructures}).
In these cases, physical quantities, including currents, are expressed in terms of certain hypergeometric series. 
Based on the exact analysis, we investigate the effects of the network geometries on transport properties.
In Sec. \ref{sec:conclusion}, we summarize our results.

\section{Model}
\label{sec:model}
\begin{figure*}[tbh]
    \centering
    \includegraphics[height=10cm]{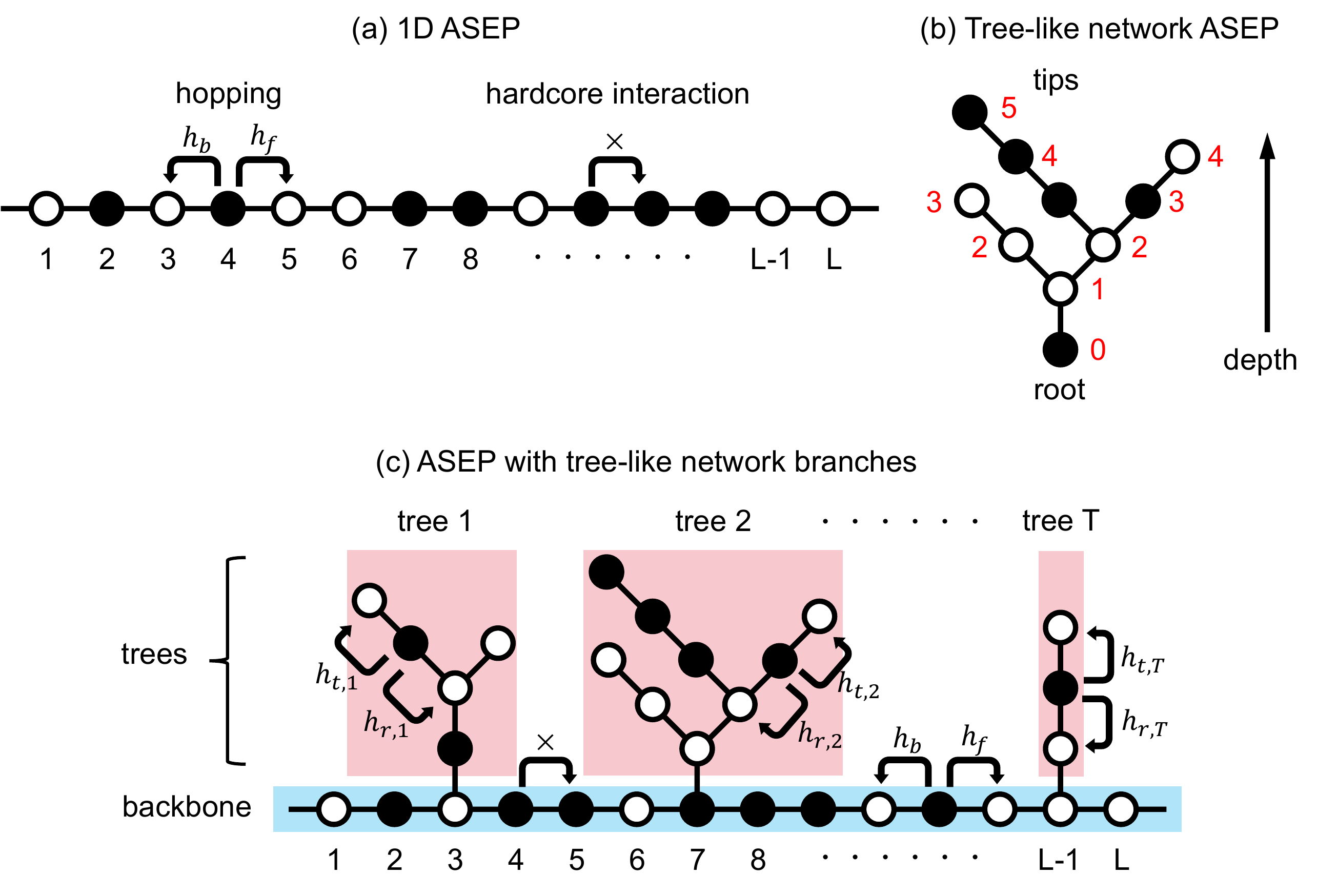}
    \caption{Schematic diagrams of the models. White (black) dots represent empty (occupied) sites. (a) 1D ASEP. Particles hop to the right (left) neighboring site at rate $h_{f}$ ($h_{b}$) if the target site is empty.  (b) Tree-like network ASEP. Particles hop toward tips (root) at rate $h_{t}$ ($h_{r}$) if the target site is empty. The red numbers represent the depth of the site, which is defined as the distance between the site and the root. (c) ASEP with tree-like network branches. Particles on the backbone hop to the right (left) at rate $h_{f}$ ($h_{b}$) if the target site is empty. Particles on the $i$-th tree hop toward tips (root) at rate $h_{t,i}$ ($h_{r,i}$) if the target site is empty.}
    \label{fig:asep_fig}
\end{figure*}

The ASEP is a continuous-time Markov process that describes the biased random walks of many particles with hard-core interactions on a 1D lattice (Fig. \ref{fig:asep_fig}(a)). The updating rule is as follows. Each particle moves to the right (left) neighboring site at rate $h_f$ ($h_b$) if the target site is empty. Due to hard-core interactions, each site can contain at most one particle.

Motivated by modeling proton transport along oxygen networks in proton-conducting solid oxides, this work extends the ASEP to systems that possess tree-like network branches (Fig. \ref{fig:asep_fig}(c)). 
In solid oxides, protons are considered to migrate primarily along oxygen networks. Since each oxygen can typically bind at most one proton, proton transport can be modeled as an exclusion process by regarding oxygens as sites and protons as particles. Although the geometry of oxygen networks can vary depending on the material, we focus here on 1D pathways with branching structures and introduce the following model.

We consider a system on a 1D lattice with tree-like network branches (Fig. \ref{fig:asep_fig}(c)). The tree-like network branches contain no closed loops. We refer to the main 1D lattice as the backbone, and to the tree-like network branches as trees. 
On the backbone, each particle hops to the right (left) neighboring site at rate $h_f$ ($h_b$) if the target site is empty. 
We consider periodic boundary conditions along the backbone direction; a particle reaching one end of the backbone appears at the opposite end.
On the $i$-th tree ($i=1,2,\cdots,T$), each particle hops towards the tips (root) at rate $h_{t,i}$ ($h_{r,i}$) if the target site is empty. We impose $h_{t,i}\neq0$ and $h_{t,r}\neq0$ for simplicity. Here, the term root refers to the junction between the backbone and the tree, and the term tips denotes the terminal ends of the trees (Fig. \ref{fig:asep_fig}(b)).
Closed boundary conditions are imposed along the tree directions; particles are not allowed to hop beyond the tips of the tree.

The time evolution of this model is described by the master equation. 
We denote by $P(C,t)$ the probability of the system being in a configuration $C$ at time $t$.
The master equation is given by
\begin{align}
\begin{split}
    &\frac{d}{dt}P(C,t) \\
    &= \sum_{C' \neq C} \left[ W(C' \to C) P(C',t) - W(C \to C')P(C,t) \right]
\end{split}
\label{eq:master_eq}
\end{align}
where $W(C \to C')$ denotes a transition rate from a configuration $C$ to $C'$.

Since the number of particles $N$ is conserved during the time evolution, a unique steady state exists for each $N$. 
The steationary distribution  $P_{\text{st}}(C)$ is given by the solution of the master equation under stationary condition:
\begin{align}
    \frac{d}{dt}P(C,t) =0.
\end{align}
Therefore, the master equation for the steady state is 
\begin{align}
    \sum_{C' \neq C} \left[ W(C' \to C) P_{\text{st}}(C) - W(C \to C')P_{\text{st}}(C) \right]=0.
    \label{eq:steady_master_eq}
\end{align}
In the long-time limit, any initial state relaxes to the steady state.
The present work focuses on the steady-state properties of this model.

\section{Steady state}
\label{sec:steady}
In this section, we present the exact stationary distribution of the ASEP with tree-like network branches.
The strategy for the construction of the steady state is the transition decomposition \cite{PhysRevResearch.6.033030}; decompose the complex ASEP into a combination of simple ASEPs. 
The present model  (Fig. \ref{fig:asep_fig}(c)) can be decomposed into a combination of a periodic 1D ASEP (Fig. \ref{fig:asep_fig}(a)) and closed ASEPs on trees, each of which is referred to as a tree-like network ASEP (Fig. \ref{fig:asep_fig}(b)).
Therefore, before discussing the ASEP with tree-like network branches, we first consider the case of the periodic 1D ASEP and the closed tree ASEP.

\subsection{Periodic 1D ASEP}
We consider the 1D ASEP in periodic boundary conditions (Fig. \ref{fig:asep_fig}(a)). Particles hop to the right (left) neighboring site at rate $h_{f}$ ($h_{b}$) if the target site is empty.
The stationary distribution in this situation is a well-known result \cite{blythe2007nonequilibrium};
all configurations are realized with equal probability. The stationary distribution $P_{\text{st,p}}(C)$ for a system of $N$ particles on $L$ sites is given by
\begin{align}
    P_{\text{st,p}}(C)=\frac{1}{Z_{\text{p},N}}, \hspace{20pt} Z_{\text{p},N} :=  \binom{L}{N},
    \label{eq:1dasep-stationary-distribution}
\end{align}
where $Z_{\text{p},N}$ is the normalization constant to satisfy $\sum_{C}P(C)=1$. 

This is confirmed as follows. We denote by $N_c(C)$ the number of clusters in a configuration $C$.
Transitions occur only at the boundaries of particle clusters; at the right (left) boundary, a particle hops to the right (left) with rate $h_{f}$ ($h_{b}$).
Under periodic boundary conditions, a particle at the boundary of clusters can always hop. Therefore, the number of configurations that can be reached by a rightward hop is equal to the number reached by a leftward hop, and both are equal to $N_c(C)$. Accordingly,
\begin{align}
\begin{split}    
    \sum_{C'\neq C} W_{\text{p}}(C \to C') &= \sum_{C' \in \bm{C}_{f}} h_{f} + \sum_{C' \in \bm{C}_{b}} h_{b} \\
    &= (h_{f}+h_{b})N_c(C),
\end{split}
\label{eq:trans_rel}
\end{align}
where $\bm{C}_{f}$ ($\bm{C}_{b}$) denotes a set of configurations that can be reached from $C$ by a rightword (leftword) hop, and $W_{\text{p}}(C \to C')$ represents a transition rate of the periodic 1D ASEP.
In addition, 
\begin{align}
\begin{split}    
    \sum_{C'\neq C} W_{\text{p}}(C' \to C) &= \sum_{C' \in \bm{C}_{f}} h_{b} + \sum_{C' \in \bm{C}_{b}} h_{f} \\
    &= (h_{f}+h_{b})N_c(C).
\end{split}
\label{eq:reverse_trans_rel}
\end{align}
From Eqs. (\ref{eq:trans_rel}) and (\ref{eq:reverse_trans_rel}), we obtain the following relation:
\begin{align}
    \sum_{C'\neq C} W_{\text{p}}(C' \to C) = \sum_{C'\neq C} W_{\text{p}}(C \to C')
    \label{eq:1dasep_transition_relation}
\end{align}
Then, from Eqs. (\ref{eq:1dasep-stationary-distribution}) and (\ref{eq:1dasep_transition_relation}), 
\begin{align}
    \begin{split}
        &\sum_{C' \neq C} \left[ W_{\text{p}}(C' \to C) P_{\text{st,p}}(C) - W_{\text{p}}(C \to C')P_{\text{st,p}}(C) \right]\\
        &\hspace{20pt}=\frac{1}{Z_{\text{p},N}}\left[ W_{\text{p}}(C' \to C)  - W_{\text{p}}(C \to C') \right]\\
        &\hspace{20pt}=0.
    \end{split}
\end{align}
Therefore, Eq. (\ref{eq:1dasep-stationary-distribution}) is the solution of the stationary master equation Eq. (\ref{eq:steady_master_eq}).

\subsection{Closed tree-like network ASEP}
We consider the tree-like network ASEP in closed boundary conditions (Fig. \ref{fig:asep_fig}(b)). Particles hop toward tips (root) at rate $h_{t}$ ($h_{r}$) if the target site is empty.
We define the depth $d$ at the site as the distance between the site and the root. As shown in Fig. \ref{fig:asep_fig}(b), the distance $d$ counts the minimum number of hops required to reach the site from the root. 

We found that the stationary distribution $P_{\text{st,t}}(C)$ for a system of $N$ particles on $L$ sites is given by
\begin{align}
\begin{split}    
    &P_{\text{st,t}}(C)= \frac{1}{Z_{\text{t},N}} q^{\sum_{i=1}^N d_i(C)} \\
    &Z_{\text{t},N} =\sum_{C} q^{\sum_{i=1}^N d_i(C)},
    \end{split}
    \label{eq:closed-tree-steady-dist}
\end{align}
where $d_i(C)$ is the depth of the site where the $i$-th particle ($i=1,2,\cdots$,N) exists for a configuration $C$, and 
\begin{align}
    q:=\frac{h_{t}}{h_{r}}.
\end{align}

This is proved as follows. 
Transitions can be classified into two types: those caused by particle hopping toward the tips, and those caused by hopping toward the roots.
Accordingly, the master equation (\ref{eq:master_eq}) is written as 
\begin{align}
\begin{split}
    &\frac{d}{dt}P_{\text{t}}(C,t) \\
    &= \sum_{C' \in \bm{C}_{t}} \left[ h_{r} P_{\text{t}}(C',t) -h_{t} P_{\text{t}}(C,t) \right] \\
    &+\sum_{C' \in \bm{C}_{r}} \left[h_{t} P_{\text{t}}(C',t) - h_{\text{r}}P_{\text{t}}(C,t) \right],
\end{split}
\label{eq:tree_asep_master_eq}
\end{align}
where $\bm{C}_{t}$ ($\bm{C}_{r}$) represents a set of configurations that can be reached from $C$ by a hopping toward the tips (root).
In addition, 
\begin{align}
    \sum_{i=1}^{N} d_{i}(C') =  
    \begin{dcases}
    \sum_{i=1}^N d_{i}(C) +1 & \hspace{5pt} \text{for} \hspace{5pt} C'\in \bm{C}_{t} \\
    \sum_{i=1}^N d_{i}(C) -1 &  \hspace{5pt} \text{for} \hspace{5pt} C'\in \bm{C}_{r}
    \end{dcases}
    \label{eq:depth-transition}
\end{align}
is satisfied. 
From Eqs. (\ref{eq:closed-tree-steady-dist}) and (\ref{eq:depth-transition}),
we find
\begin{align}
\begin{dcases}
     h_{r} P_{\text{st,t}}(C') -h_{t} P_{\text{st,t}}(C) =0 & \hspace{5pt} \text{for} \hspace{5pt} C'\in \bm{C}_{t} \\
    h_{t} P_{\text{st,t}}(C') - h_{\text{r}}P_{\text{st,t}}(C) =0 & \hspace{5pt} \text{for} \hspace{5pt} C'\in \bm{C}_{r},
\end{dcases}
\end{align}
and the right-hand side of the master equation (\ref{eq:tree_asep_master_eq}) is
\begin{align}
    \begin{split}
    &\sum_{C' \in \bm{C}_{t}} \left[ h_{r} P_{\text{st,t}}(C') -h_{t} P_{\text{st,t}}(C) \right] \\
    &\hspace{20pt}+\sum_{C' \in \bm{C}_{r}} \left[h_{t} P_{\text{st,t}}(C') - h_{\text{r}}P_{\text{st,t}}(C) \right] =0.
\end{split}
\end{align}
Therefore, we can confirm that Eq. (\ref{eq:closed-tree-steady-dist}) is the stationary distribution of the closed tree ASEP.

\subsection{ASEP with tree-like network branches}
We consider the ASEP with tree-like network branches (Fig. \ref{fig:asep_fig}(c)), which is the main subject of this work. 
On the backbone, particles hop toward the right (left) at rate $h_{f}$ ($h_{b}$) if the target site is empty. 
On the $i$-th tree ($i=1,2,\cdots,T$), particles hop toward the tips (root) at rate $h_{t,i}$ ($h_{r,i}$) if the target site is empty. 

The stationary distribution $P_{\text{st,s}}(C)$ for a system of $N$ particles on $L$ sites is given by
\begin{align}
\begin{split}    
    &P_{\text{st,s}}(C)=\frac{1}{Z_{\text{s},N}} \prod_{i=1}^{T} q_{i}^{\sum_{j=1}^{n_i(C)} d_{i,j}(C)} \\
    &Z_{\text{s},N}=\sum_{C} \prod_{i=1}^{T} q_{i}^{\sum_{j=1}^{n_i(C)} d_{i,j}(C)},
    \end{split}
    \label{eq:st-dist_asep-with-treesub}
\end{align}
where $T$ denotes the number of trees, $n_i(C)$ represents the number of particles on $i$-th tree, $d_{i,j}(C)$ is the depth of the site where the $j$-th particle ($j=1,2,\cdots,n_i$) on $i$-th tree exists, and $q_i$ is defined as
\begin{align}
    q_i := \frac{h_{t,i}}{h_{r,i}}.
\end{align}

This is proved as follows. We decompose the ASEP with tree-like network branches into the periodic 1D ASEP and the closed tree-like network ASEPs; the master equation (\ref{eq:master_eq}) is expressed as follows:
\begin{align}
\begin{split}
    &\frac{d}{dt}P_{\text{s}}(C,t) \\
    &= \sum_{C' \in \bm{B}} \left[ W(C' \to C) P_{\text{s}}(C',t) - W(C \to C')P_{\text{s}}(C,t) \right] + \\
    &\sum_{i=1}^{T}\sum_{C' \in \bm{T_{i}}} \left[ W(C' \to C) P_{\text{s}}(C',t) - W(C \to C')P_{\text{s}}(C,t) \right]
\end{split}
\label{eq:master_ASEP_with_substructure}
\end{align}
where $\bm{B}$ ($\bm{T_i}$) represents a set of configurations that can be reached from the configuration $C$ by the transition on the backbone ($i$-th tree).
Obviously, $\bm{B}$ ($\bm{T}$) is equivalent to the set of configurations that can be reached by the transition of the periodic 1D ASEP (the closed tree ASEP).
Therefore, the following relations are satisfied:
\begin{align}
    \begin{split}
        &\sum_{C' \in \bm{B}} \left[ W(C' \to C)  - W(C \to C') \right]=0 \\
        &\sum_{C' \in \bm{T_{i}}} \left[ W(C' \to C) q_i^{\sum_{j=1}^{n_i} d_{i,j}(C')} \right. \\
        &\hspace{25pt} \left. - W(C \to C') q_i^{\sum_{j=1}^{n_i} d_{i,j}(C)} \right] = 0.
    \end{split}
    \label{eq:dec_rel}
\end{align}
From Eqs. (\ref{eq:st-dist_asep-with-treesub}) and (\ref{eq:dec_rel}), the right-hand side of the master equation (\ref{eq:master_ASEP_with_substructure}) is 
\begin{align}
    \begin{split}
    &\sum_{C' \in \bm{B}} \left[ W(C' \to C) P_{\text{st,s}}(C') - W(C \to C')P_{\text{s,t}}(C) \right] +\\
    &\sum_{i=1}^{T}\sum_{C' \in \bm{T_{i}}} \left[ W(C' \to C) P_{\text{st,s}}(C') - W(C \to C')P_{\text{st,s}}(C) \right] \\
    &=\frac{1}{Z_{\text{s},N}} \prod_{i=1}^{T} q_{i}^{\sum_{j=1}^{n_i(C)} d_{i,j}(C)} \times \\
    &\hspace{35pt} \sum_{C'\in \bm{B}}\left[ W(C' \to C)  - W(C \to C') \right] \\
    &+\frac{1}{Z_{\text{s},N}} \sum_{i=1}^{T} \prod_{\ell \neq i} q_{l}^{\sum_{j=1}^{n_{\ell}} d_{\ell,j}(C)} \times \\
    & \hspace{35pt} \sum_{C' \in \bm{T}_i} \left[ W(C' \to C) q_i^{\sum_{j=1}^{n_i} d_{i,j}(C')} \right. \\
    &\hspace{60pt} \left. - W(C \to C') q_i^{\sum_{j=1}^{n_i} d_{i,j}(C)} \right]\\ 
        &= 0.
    \end{split}
\end{align}
Accordingly, Eq. (\ref{eq:st-dist_asep-with-treesub}) provides the exact stationary distribution of the AESP with tree-like network branches.

\section{transport properties}
\label{sec:transport}
In the previous section, we obtained the stationary distribution of the ASEP with tree-like network branches (\ref{eq:st-dist_asep-with-treesub}). 
Based on the exact solutions, we investigate the transport properties. 
In particular, we elucidate the effect of the network structure on the current.

\begin{figure}[tbh]
    \centering
    \includegraphics[height=8.5cm]{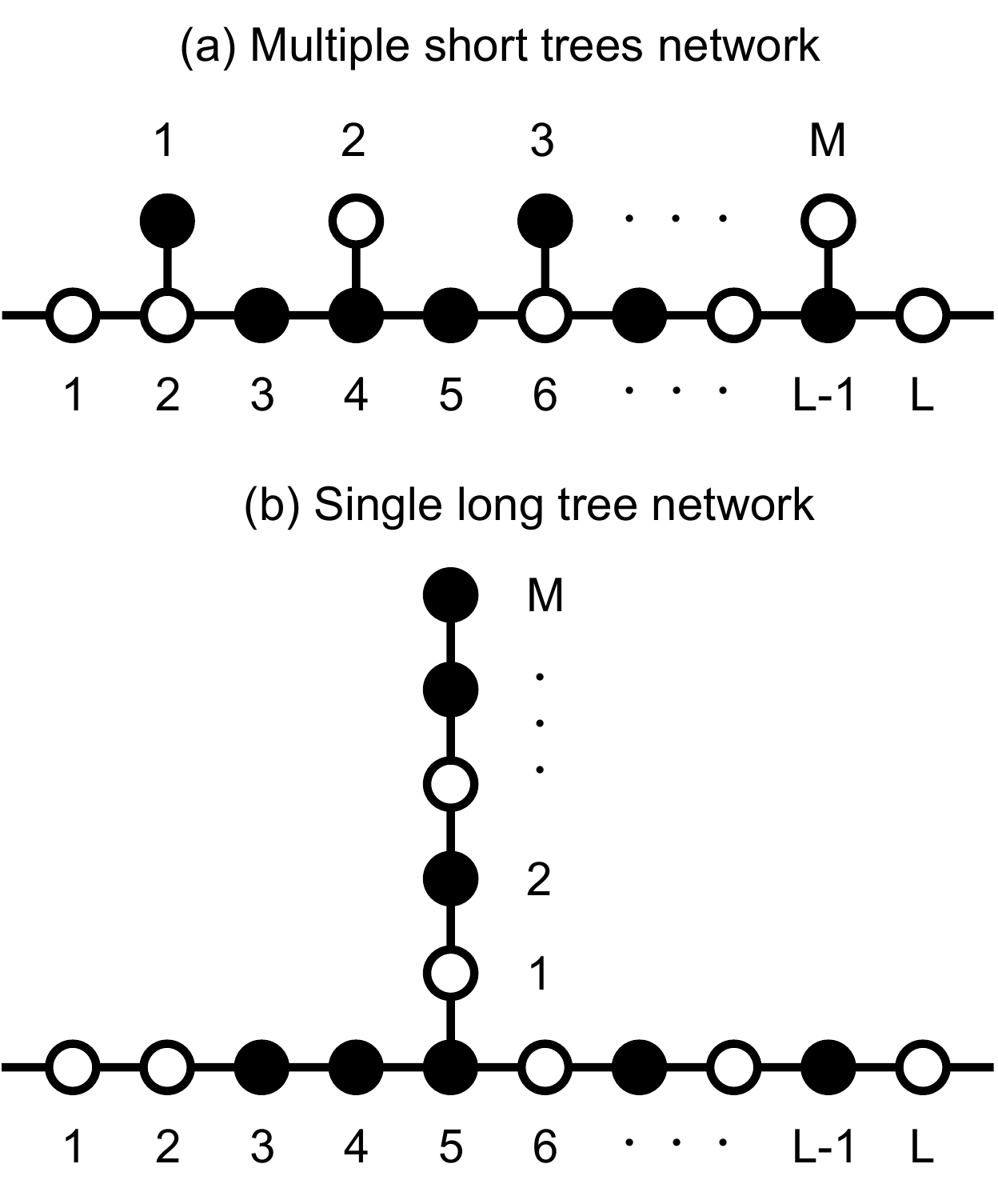}
    \caption{Schematic diagrams of the tree substructures. White (black) dots represent empty (occupied) sites. (a) Multiple short trees network. The number of trees is $M$, and the maximum depth of each tree is one. (b) Single long tree network. The number of trees is one, and the maximum depth of the tree is $M$.}
    \label{fig:substructures}
\end{figure}

As fundamental examples, we examine two types of tree-like networks: (i) a multiple short trees network (Fig. \ref{fig:substructures}(a)) and (ii) a single long tree network (Fig. \ref{fig:substructures}(b)).
We consider a system where the number of sites on the backbone is $L$, and the number of sites on the trees is $M$.
A multiple short trees network contains $M$ trees with depth one, and a single long tree network has a tree with depth $M$.

\subsection{Multiple short trees network}
We consider the ASEP with tree-like network branches that have multiple short trees (Fig. \ref{fig:substructures}(a)). 
The number of trees is $M$, and the maximum depth of each tree is one.
For simplicity, we assume that the hopping rates on trees are homogeneous: $h_{t,i}=h_{t}$, $h_{r,i}=h_{r}$ ($i=1,2,\cdots,M$).

To describe a configuration of a system, we introduce variables $n_j$ and $\tau_j$.
Suppose that, among the $N$ particles in the system, $i$ particles are located in trees and $N-i$ particles are in the backbone ($i=0,1,\cdots, M$).
$n_j$ describes the position of the $j$-th particle ($j=1,2,\cdots,N-i$) on the backbone. $\tau_i$ denotes the position of the $j$-th particle ($j=1,2,\cdots,i$) on trees.
A configuration of a system is described by a tuple of variables $\{n|\tau\}_i:= \{n_1, n_2,\cdots,n_{N-i}| \tau_1,\tau_2,\cdots,\tau_i \}$ ($i=0,1,\cdots, M$). 

From the exact solution of the ASEP with tree substratures (\ref{eq:st-dist_asep-with-treesub}), the stationary distribution of the model $P_{\text{st,ms}}(\{n|\tau\}_i)$ is given by
\begin{align}
    P_{\text{st,ms}}(\{n|\tau\}_i) = \frac{q^{i}}{Z_{\text{st,ms}}},
    \label{eq:st_dist_maltiple_short_tree}
\end{align}
where
\begin{align}
\begin{split}
    Z_{\text{st,ms}} &=\sum_{i=0}^{M} \sum_{\{n|\tau\}_i} q^{i} \\
    &= \sum_{i=0}^{M} \binom{L}{N-i} \binom{M}{i} q^i. 
\end{split}
\label{eq:pf_maltiple_short_tree}
\end{align}
$Z_{\text{st,ms}}$ is expressed in terms of the hypergeometric series  ${}_2 F_{1}(a,b,c;z)$:
\begin{align}
    {}_2 F_{1}(a,b,c;z) := \sum_{i=0}^{\infty} \frac{(a)_i (b)_i}{(c)_i} \frac{z^i}{i!},
\end{align}
where we use the Pochhammer symbol:
\begin{align}
    (a)_i := \prod_{k=0}^{i-1}(a+k).
\end{align}
From the following relations:
\begin{align}
    \begin{split}
        \binom{L}{N-i}&=(-1)^i \binom{L}{N}\frac{(-N)_i}{(L-N+1)_i} \\
        \binom{M}{i} &= (-1)^i \frac{(-M)_i}{i!},
    \end{split}
    \label{eq:derivation_hyp}
\end{align}
Eq. (\ref{eq:pf_maltiple_short_tree}) is given by
\begin{align}
    Z_{\text{st,ms}}=\binom{L}{N} {}_2 F_{1}(-M,-N,L-N+1;q).
    \label{eq:pf_hyper_multiple_short}
\end{align}

We evaluate the particle density $\rho_{\text{b}}$ and current $j_{\text{b}}$ in the backbone.
We denote the expectation value of a physical quantity $A(C)$ in the stationary state as $\bra A \ket := \sum_{C} P_{\text{st}}(C) A(C)$.
Then, we define $\rho_{\text{b}}$ and $j_{\text{b}}$ as
\begin{align}
\begin{split}    
    \rho_{\text{b}}&:=\bra m_i \ket    \\
    j_{\text{b}}&:= h_{f} \bra m_i (1-m_{i+1})\ket -h_{b} \bra (1-m_i)m_{i+1}\ket,
    \label{eq:physical_q_bb}
\end{split}
\end{align}
where $m_i$ is a variable that describes the state of site $i$ in the backbone; if site $i$ is empty (ocuppied), $m_i=0$ ($m_i=1$).

From Eqs. (\ref{eq:st_dist_maltiple_short_tree}) and (\ref{eq:physical_q_bb}), the particle density and the current in the backbone with multiple short trees are given by
\begin{align}
\begin{split}    
    \rho_{\text{b,ms}} &= \frac{\sum_{i=0}^{M}\binom{L-1}{N-i-1}\binom{M}{i}q^{i}}{\sum_{i=0}^{M}\binom{L}{N-i}\binom{M}{i}q^{i}} \\
    &=\frac{N}{L} \frac{{}_2 F_1(-M,-N+1,L-N+1;q)}{{}_2 F_1(-M,-N,L-N+1;q)},
\end{split}
\label{eq:multiple_short_density}
\end{align}
\begin{align}
    \begin{split}
        j_{\text{b,ms}} &= (h_f-f_b)\frac{\sum_{i=0}^{M}\binom{L-2}{N-i-1}\binom{M}{i}q^{i}}{\sum_{i=0}^{M}\binom{L}{N-i}\binom{M}{i}q^{i}} \\
        &= (h_f-f_b)\frac{N(L-N)}{L(L-1)} \frac{{}_2 F_1(-M,-N+1,L-N;q)}{{}_2 F_1(-M,-N,L-N+1;q)}.
    \end{split}
\label{eq:multiple_short_current}
\end{align}

To understand the effect of the tree-like network branches, we compare the periodic 1D ASEP case. 
The particle density and the current of the periodic 1D ASEP for a $N$ particle system with $L$ sites are known as 
\begin{align}
    \begin{split}
        \rho_{\text{p}}&=\frac{N}{L} \\
        j_{\text{p}}&= (h_f-f_b)\frac{N(L-N)}{L(L-1)}.
    \end{split}
    \label{eq:pbc_asep_phys_quantities}
\end{align}
Therefore, the effect of the multiple short trees network is expressed in terms of the hypergeometric series as follows:
\begin{align}
    \begin{split}
    \frac{\rho_{\text{b,ms}}}{\rho_{\text{p}}}&=\frac{{}_2 F_1(-M,-N+1,L-N+1;q)}{{}_2 F_1(-M,-N,L-N+1;q)}, \\
        \frac{j_{\text{b,ms}}}{j_{\text{p}}}&= \frac{{}_2 F_1(-M,-N+1,L-N;q)}{{}_2 F_1(-M,-N,L-N+1;q)}.
    \end{split}
    \label{eq:ms_tree_against_pbc_asap}
\end{align}

\begin{figure}[tbh]
    \centering
    \includegraphics[height=10cm]{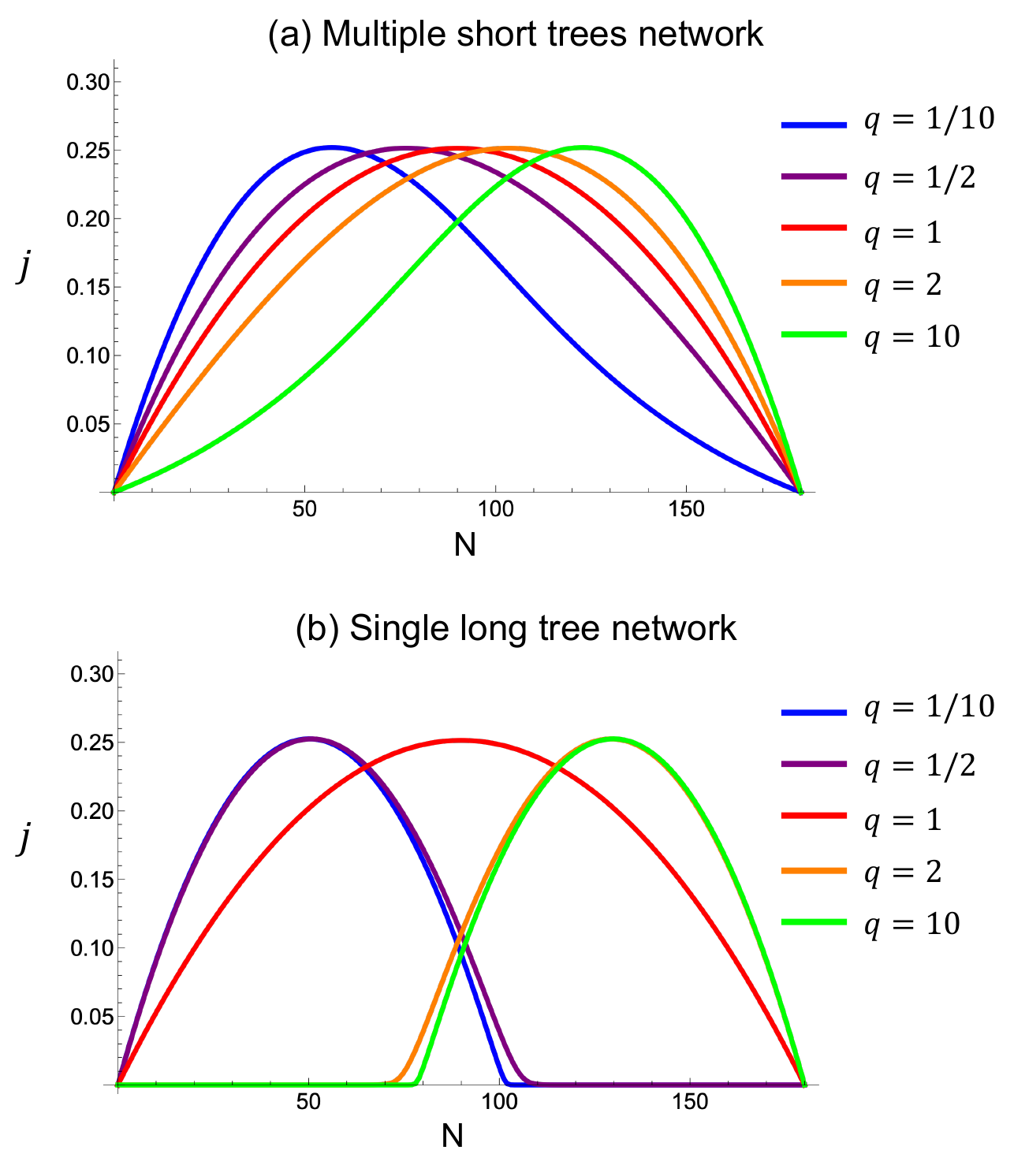}
    \caption{Currents of the backbone against the number of particles $N$ with (a) multiple short trees network $j_{\text{b,ms}}$ (Eq. (\ref{eq:multiple_short_current})) and (b) single long tree network ($L=100$, $M=80$, and $(h_f,h_b)=(1,0)$).}
    \label{fig:current_tree}
\end{figure}

In Fig. \ref{fig:current_tree}(a), we show the current $j_{\text{b,ms}}$ (Eq. (\ref{eq:multiple_short_current})) against the number of partilce $N$. 
When $q=1$, the graph of the current is symmetric against $N$, which is equivalent to the periodic 1D ASEP case. In contrast, in the case of $q \neq 1$, the graph becomes asymmetrically skewed, and the peak position shifts away from the center. When $q <1$ ($q>1$), the peak shifts towards the low-density (high-density) region.

\subsection{Single long tree network}
We consider the ASEP with tree-like network branches that have a single short tree (Fig. \ref{fig:substructures}(b)). 
The number of trees is one, and the maximum depth of the tree is $M$.

To describe a configuration of a system, we introduce variables $n_j$ and $\tau_j$.
Suppose that, among the $N$ particles in the system, $i$ particles are located in the tree and $N-i$ particles are in the backbone ($i=0,1,\cdots, M$).
$n_j$ describes the position of the $j$-th particle ($j=1,2,\cdots,N-i$) on the backbone. $\tau_i$ denotes the depth of the $j$-th particle ($j=1,2,\cdots,i$) on the tree.
A configuration of a system is described by a set of variables $\{n|\tau\}_i:= \{n_1, n_2,\cdots,n_{N-i}| \tau_1,\tau_2,\cdots,\tau_i \}$ ($i=0,1,\cdots, M$). 

From the exact solution of the ASEP with tree substratures (\ref{eq:st-dist_asep-with-treesub}), the stationary distribution of the model $P_{\text{st,sl}}(\{n|\tau\}_i)$ is given by
\begin{align}
    P_{\text{st,sl}}(\{n|\tau\}_i) = \frac{1}{Z_{\text{st,sl}}}q^{\sum_{j=1}^{i}\tau_j},
    \label{eq:st_dist_single_long_tree}
\end{align}
where
\begin{align}
\begin{split}
    Z_{\text{st,sl}} &=\sum_{i=0}^{M} \sum_{\{n|\tau\}_i} q^{\sum_{j=1}^{i}\tau_j} \\
    &= \sum_{i=0}^{M} \binom{L}{N-i} 
    \binom{M}{i}_q
     q^{\frac{i(i+1)}{2}}.
\end{split}
\label{eq:pf_single_long_tree}
\end{align}
Here $\binom{m}{r}_q$ denotes the $q$-binomial:
\begin{align}
    \binom{m}{r}_q:= \frac{[m]_q}{[r]_q [m-r]_q},
\end{align}
where $[n]_q$ is the $q$-integer:
\begin{align}
    [n]_q =\sum_{i=0}^{n-1} q^{i}.
\end{align}
In the manipulation of Eq. (\ref{eq:pf_single_long_tree}), we use the following relation:
\begin{align}
    \sum_{1 \le \tau_1 < \cdots < \tau_i \le M} q^{\sum_{j=1}^i \tau_j} = \binom{M}{i}_q q^{\frac{i(i+1)}{2}}.
\end{align}
$ Z_{\text{st,sl}}$ (\ref{eq:pf_single_long_tree}) is expressed in terms of the partially q-deformed hypergeometric series ${}_2 \phi_{1}(a,b,c;z)$ (mixed hypergeometric series \cite{mixed}) that is defined as 
\begin{align}
    {}_2 \phi_{1}(a,b,c;z) := \sum_{i=0}^{\infty} \frac{(q^a;q)_i (b)_i}{(c)_i} \frac{z^i}{(q;q)_i},
\end{align}
where we use the $q$-Pochhammer symbol:
\begin{align}
    (a;q)_i := \prod_{k=0}^{i-1} (1-aq^k).
\end{align}
From Eq. (\ref{eq:derivation_hyp}) and the following relation:
\begin{align}
    \binom{M}{i}_q q^{\frac{i(i+1)}{2}} = (-q)^{(M+1)i} \frac{(q^{-M};q)}{(q;q)_i},
\end{align}
Eq. (\ref{eq:pf_single_long_tree}) is given by
\begin{align}
    Z_{\text{st,sl}}= \binom{L}{N} {}_2 \phi_{1}(-M,-N,L-N+1;q^{M+1}).
    \label{eq:pf_hyper_single_long_tree}
\end{align}

From Eqs. (\ref{eq:physical_q_bb}) and (\ref{eq:st_dist_single_long_tree}), the particle density $\rho_{\text{b,sl}}$ and the current $j_{\text{b,sl}}$ in the backbone with the single long tree are given by
\begin{align}
\begin{split}    
    \rho_{\text{b,sl}} &= \frac{\sum_{i=0}^{M}\binom{L-1}{N-i-1}\binom{M}{i}_q q^{\frac{i(i+1)}{2}}}{\sum_{i=0}^{M}\binom{L}{N-i}\binom{M}{i}_q q^{\frac{i(i+1)}{2}}} \\
    &=\frac{N}{L} \frac{{}_2 \phi_1(-M,-N+1,L-N+1;q^{M+1})}{{}_2 \phi_1(-M,-N,L-N+1;q^{M+1})},
\end{split}
\label{eq:single_long_density}
\end{align}
\begin{align}
    \begin{split}
        j_{\text{b,sl}} &= (h_f-f_b)\frac{\sum_{i=0}^{M}\binom{L-2}{N-i-1}\binom{M}{i}_q q^{\frac{i(i+1)}{2}}}{\sum_{i=0}^{M}\binom{L}{N-i}\binom{M}{i}_q q^{\frac{i(i+1)}{2}}} \\
        &= (h_f-f_b)\frac{N(L-N)}{L(L-1)} \frac{{}_2 \phi_1(-M,-N+1,L-N;q^{M+1})}{{}_2 \phi_1(-M,-N,L-N+1;q^{M+1})}.
    \end{split}
\label{eq:single_long_current}
\end{align}

To understand the effect of the single long tree network, we compare the periodic 1D ASEP case. 
From Eqs. (\ref{eq:pbc_asep_phys_quantities}), (\ref{eq:single_long_density}), and (\ref{eq:single_long_current}),
\begin{align}
    \begin{split}
    \frac{\rho_{\text{b,sl}}}{\rho_{\text{p}}} &=  \frac{{}_2 \phi_1(-M,-N+1,L-N+1;q^{M+1})}{{}_2 \phi_1(-M,-N,L-N+1;q^{M+1})}\\
        \frac{j_{\text{b,sl}}}{j_{\text{p}}}&= \frac{{}_2 \phi_1(-M,-N+1,L-N;q^{M+1})}{{}_2 \phi_1(-M,-N,L-N+1;q^{M+1})}.
    \end{split}
\end{align}
Therefore, the effect of the single long tree structure is expressed in terms of the mixed hypergeometric series.

In Fig. \ref{fig:current_tree}(b), we show the current $j_{\text{b,ms}}$ (Eq. (\ref{eq:single_long_current})) against the number of partilce $N$. 
When $q=1$, the graph of the current is symmetric against $N$, which is equivalent to the periodic 1D ASEP case. On the other hand, in the case of $q \neq 1$, the graph becomes asymmetrically skewed, and the peak position shifts away from the center. 
In contrast to the case of the multiple short trees network, this case exhibits two distinct regions: one with almost no flow and the other with flow. This suggests that longer branches amplify the effect of interparticle interactions on transport.

\section{Conclusion}
\label{sec:conclusion}

In this paper, we have presented an extension of the ASEP that incorporates tree-like network branches as a model for proton transport along oxygen networks in proton-conducting solid oxides.
We have derived the exact stationary distribution of the model and investigated its transport properties based on these exact results.
As representative examples, we have considered a multiple short trees network and a single long tree network.
A comparison of these two networks reveals that their transport properties differ substantially (Fig. \ref{fig:current_tree}), indicating that longer branches enhance the impact of interparticle interactions on transport.
The difference between the networks corresponds to the difference in the hypergeometric series that describes their physical quantities. 
Those for the former are written in terms of the hypergeometric series, whereas those for the latter are expressed in terms of the mixed hypergeometric series.

While our analysis has focused on quasi-1D networks, oxygen networks in real materials can exhibit a wide variety of geometries, including two-dimensional and three-dimensional transport pathways. Exploring exact solutions of the ASEP and the associated transport properties on more complex network structures remains an important direction for future work. In addition, developing perturbative approaches around exactly solvable limits \cite{10.21468/SciPostPhys.17.3.092} offers another promising route to capture more complex and realistic transport phenomena.

\begin{acknowledgments}
This work was supported by JSPS KAKENHI Grant No. JP24K16976 and No. JP24H02203.
\end{acknowledgments}






\bibliography{apssamp}

@article{PhysRevResearch.6.033030,
  title = {Exact steady states in the asymmetric simple exclusion process beyond one dimension},
  author = {Ishiguro, Yuki and Sato, Jun},
  journal = {Phys. Rev. Res.},
  volume = {6},
  issue = {3},
  pages = {033030},
  numpages = {7},
  year = {2024},
  month = {Jul},
  publisher = {American Physical Society},
  doi = {10.1103/PhysRevResearch.6.033030},
  url = {https://link.aps.org/doi/10.1103/PhysRevResearch.6.033030}
}

@article{Ishiguro_2025,
doi = {10.1088/1751-8121/adb085},
url = {https://dx.doi.org/10.1088/1751-8121/adb085},
year = {2025},
month = {feb},
publisher = {IOP Publishing},
volume = {58},
number = {7},
pages = {075001},
author = {Ishiguro, Yuki and Sato, Jun},
title = {Exact analysis of the two-dimensional asymmetric simple exclusion process with attachment and detachment of particles},
journal = {Journal of Physics A: Mathematical and Theoretical}
}

@article{Derrida_1993,
doi = {10.1088/0305-4470/26/7/011},
url = {https://dx.doi.org/10.1088/0305-4470/26/7/011},
year = {1993},
month = {apr},
publisher = {},
volume = {26},
number = {7},
pages = {1493},
author = {B Derrida and  M R Evans and  V Hakim and  V Pasquier},
title = {Exact solution of a 1D asymmetric exclusion model using a matrix formulation},
journal = {J. Phys. A: Math. Theor.}
}

@article{Crampe_2014,
doi = {10.1088/1742-5468/2014/11/P11032},
url = {https://dx.doi.org/10.1088/1742-5468/2014/11/P11032},
year = {2014},
month = {nov},
publisher = {IOP Publishing and SISSA},
number = {11},
pages = {P11032},
author = {N Crampe and E Ragoucy and M Vanicat},
title = {Integrable approach to simple exclusion processes with boundaries. Review and progress},
journal = {J. Phys. Mech.: Theory Exp.}
}

@article{derrida1998exactly,
  title={An exactly soluble non-equilibrium system: the asymmetric simple exclusion process},
  author={Derrida, Bernard},
  journal={Phys. Rep.},
  volume={301},
  number={1-3},
  pages={65--83},
  year={1998},
  publisher={Elsevier}
}

@article{golinelli2006asymmetric,
  title={The asymmetric simple exclusion process: an integrable model for non-equilibrium statistical mechanics},
  author={Golinelli, Olivier and Mallick, Kirone},
  journal={J. Phys. A: Math. Gen.},
  volume={39},
  number={41},
  pages={12679},
  year={2006},
  publisher={IOP Publishing}
}

@article{blythe2007nonequilibrium,
  title={Nonequilibrium steady states of matrix-product form: a solver's guide},
  author={Blythe, Richard A and Evans, Martin R},
  journal={J. Phys. A: Math. Theor.},
  volume={40},
  number={46},
  pages={R333},
  year={2007},
  publisher={IOP Publishing}
}

@article{essler1996representations,
  title={Representations of the quadratic algebra and partially asymmetric diffusion with open boundaries},
  author={Essler, Fabian HL and Rittenberg, Vladimir},
  journal={J. Phys. A: Math. Gen.},
  volume={29},
  number={13},
  pages={3375},
  year={1996},
  publisher={IOP Publishing}
}

@article{gwa1992bethe,
  title={Bethe solution for the dynamical-scaling exponent of the noisy {B}urgers equation},
  author={Gwa, Leh-Hun and Spohn, Herbert},
  journal={Phys. Rev. A},
  volume={46},
  number={2},
  pages={844},
  year={1992},
  publisher={APS}
}

@article{kim1995bethe,
  title={Bethe ansatz solution for crossover scaling functions of the asymmetric {XXZ} chain and the {K}ardar-{P}arisi-{Z}hang-type growth model},
  author={Kim, Doochul},
  journal={Phys. Rev. E},
  volume={52},
  number={4},
  pages={3512},
  year={1995},
  publisher={APS}
}

@article{Golinelli_2004,
doi = {10.1088/0305-4470/37/10/001},
url = {https://dx.doi.org/10.1088/0305-4470/37/10/001},
year = {2004},
publisher = {},
volume = {37},
number = {10},
pages = {3321},
author = {O Golinelli and  K Mallick},
title = {Bethe ansatz calculation of the spectral gap of the asymmetric exclusion process},
journal = {J. Phys. A: Math. Gen.}
}

@article{Golinelli_2005,
doi = {10.1088/0305-4470/38/7/001},
url = {https://dx.doi.org/10.1088/0305-4470/38/7/001},
year = {2005},
publisher = {},
volume = {38},
number = {7},
pages = {1419},
author = {O Golinelli and K Mallick},
title = {Spectral gap of the totally asymmetric exclusion process at arbitrary filling},
journal = {J. Phys. A: Math. Gen.}
}

@article{PhysRevE.85.042105,
  title = {Exact relaxation dynamics in the totally asymmetric simple exclusion process},
  author = {Motegi, Kohei and Sakai, Kazumitsu and Sato, Jun},
  journal = {Phys. Rev. E},
  volume = {85},
  issue = {4},
  pages = {042105},
  numpages = {5},
  year = {2012},
  month = {Apr},
  publisher = {American Physical Society},
  doi = {10.1103/PhysRevE.85.042105},
  url = {https://link.aps.org/doi/10.1103/PhysRevE.85.042105}
}

@article{Motegi_2012,
doi = {10.1088/1751-8113/45/46/465004},
url = {https://dx.doi.org/10.1088/1751-8113/45/46/465004},
year = {2012},
month = {oct},
publisher = {IOP Publishing},
volume = {45},
number = {46},
pages = {465004},
author = {Kohei Motegi and Kazumitsu Sakai and Jun Sato},
title = {Long time asymptotics of the totally asymmetric simple exclusion process},
journal = {J. Phys. A: Math. Theor.}
}

@article{prolhac2013spectrum,
  title={Spectrum of the totally asymmetric simple exclusion process on a periodic lattice—bulk eigenvalues},
  author={Prolhac, Sylvain},
  journal={J. Phys. A: Math. Theor.},
  volume={46},
  number={41},
  pages={415001},
  year={2013},
  publisher={IOP Publishing}
}

@article{prolhac2014spectrum,
  title={Spectrum of the totally asymmetric simple exclusion process on a periodic lattice-first excited states},
  author={Prolhac, Sylvain},
  journal={J. Phys. A: Math. Theor.},
  volume={47},
  number={37},
  pages={375001},
  year={2014},
  publisher={IOP Publishing}
}

@article{prolhac2016extrapolation,
  title={Extrapolation methods and {B}ethe ansatz for the asymmetric exclusion process},
  author={Prolhac, Sylvain},
  journal={J. Phys. A: Math. Theor.},
  volume={49},
  number={45},
  pages={454002},
  year={2016},
  publisher={IOP Publishing}
}

@article{prolhac2017perturbative,
  title={Perturbative solution for the spectral gap of the weakly asymmetric exclusion process},
  author={Prolhac, Sylvain},
  journal={J. Phys. A: Math. Theor.},
  volume={50},
  number={31},
  pages={315001},
  year={2017},
  publisher={IOP Publishing}
}

@article{ishiguro2023,
  title = {Asymmetry-induced delocalization transition in the integrable non-Hermitian spin chain},
  author = {Ishiguro, Yuki and Sato, Jun and Nishinari, Katsuhiro},
  journal = {Phys. Rev. Res.},
  volume = {5},
  issue = {3},
  pages = {033102},
  numpages = {9},
  year = {2023},
  month = {Aug},
  publisher = {American Physical Society},
  doi = {10.1103/PhysRevResearch.5.033102},
  url = {https://link.aps.org/doi/10.1103/PhysRevResearch.5.033102}
}

@article{de2005bethe,
  title={Bethe ansatz solution of the asymmetric exclusion process with open boundaries},
  author={De Gier, Jan and Essler, Fabian HL},
  journal={Phys. Rev. Lett.},
  volume={95},
  number={24},
  pages={240601},
  year={2005},
  publisher={APS}
}

@article{deGier_2006,
doi = {10.1088/1742-5468/2006/12/P12011},
url = {https://dx.doi.org/10.1088/1742-5468/2006/12/P12011},
year = {2006},
publisher = {},
volume = {2006},
number = {12},
pages = {P12011},
author = {Jan de Gier and Fabian H L Essler},
title = {Exact spectral gaps of the asymmetric exclusion process with open boundaries},
journal = {J. Stat. Mech.: Theory Exp.}
}

@article{deGier_2008,
doi = {10.1088/1751-8113/41/48/485002},
url = {https://dx.doi.org/10.1088/1751-8113/41/48/485002},
year = {2008},
publisher = {},
volume = {41},
number = {48},
pages = {485002},
author = {Jan de Gier and Fabian H L Essler},
title = {Slowest relaxation mode of the partially asymmetric exclusion process with open boundaries},
journal = {J. Phys. A: Math. Theor.}
}

@article{deGier_2011,
  title = {Large {D}eviation {F}unction for the {C}urrent in the {O}pen {A}symmetric {S}imple {E}xclusion {P}rocess},
  author = {de Gier, Jan and Essler, Fabian H. L.},
  journal = {Phys. Rev. Lett.},
  volume = {107},
  issue = {1},
  pages = {010602},
  numpages = {4},
  year = {2011},
  publisher = {American Physical Society},
  doi = {10.1103/PhysRevLett.107.010602},
  url = {https://link.aps.org/doi/10.1103/PhysRevLett.107.010602}
}

@article{Wen_2015,
doi = {10.1088/0256-307X/32/5/050503},
url = {https://dx.doi.org/10.1088/0256-307X/32/5/050503},
year = {2015},
publisher = {Chinese Physical Society and IOP Publishing},
volume = {32},
number = {5},
pages = {050503},
author = {Fa-Kai Wen and Zhan-Ying Yang and Shuai Cui and Jun-Peng Cao and Wen-Li Yang},
title = {Spectrum of the {O}pen {A}symmetric {S}imple {E}xclusion {P}rocess with {A}rbitrary {B}oundary {P}arameters},
journal = {Chin. Phys. Lett.}
}

@article{Crampe_2015,
doi = {10.1088/1751-8113/48/8/08FT01},
url = {https://dx.doi.org/10.1088/1751-8113/48/8/08FT01},
year = {2015},
publisher = {IOP Publishing},
volume = {48},
number = {8},
pages = {08FT01},
author = {N Crampé},
title = {Algebraic {B}ethe ansatz for the totally asymmetric simple exclusion process with boundaries},
journal = {J. Phys. A: Math. Theor.}
}

@article{relation,
author = {Ishiguro ,Yuki and Sato ,Jun and Nishinari ,Katsuhiro},
title = {Relationships among the Asymmetric Simple Exclusion Process, the Burgers Equation and the Derivative Nonlinear Schrödinger Equation},
journal = {J. Phys. Soc. Jpn.},
volume = {90},
number = {11},
pages = {114008},
year = {2021},
doi = {10.7566/JPSJ.90.114008},
URL = {https://doi.org/10.7566/JPSJ.90.114008}}

@article{schadschneider2000statistical,
  title={Statistical physics of traffic flow},
  author={Schadschneider, Andreas},
  journal={Physica A},
  volume={285},
  number={1-2},
  pages={101--120},
  year={2000},
  publisher={Elsevier}
}

@book{schadschneider2010stochastic,
  title={Stochastic transport in complex systems: from molecules to vehicles},
  author={Schadschneider, Andreas and Chowdhury, Debashish and Nishinari, Katsuhiro},
  year={2010},
  publisher={Elsevier}
}

@article{macdonald1968kinetics,
  title={Kinetics of biopolymerization on nucleic acid templates},
  author={MacDonald, Carolyn T and Gibbs, Julian H and Pipkin, Allen C},
  journal={Biopolymers},
  volume={6},
  number={1},
  pages={1--25},
  year={1968},
  publisher={Wiley Online Library}
}

@article{klumpp2003traffic,
  title={Traffic of molecular motors through tube-like compartments},
  author={Klumpp, Stefan and Lipowsky, Reinhard},
  journal={J. Stat. Phys.},
  volume={113},
  number={1},
  pages={233--268},
  year={2003},
  publisher={Springer}
}

@article{PhysRevE.84.061141,
  title = {Exact solution of a heterogeneous multilane asymmetric simple exclusion process},
  author = {Ezaki, Takahiro and Nishinari, Katsuhiro},
  journal = {Phys. Rev. E},
  volume = {84},
  issue = {6},
  pages = {061141},
  numpages = {4},
  year = {2011},
  month = {Dec},
  publisher = {American Physical Society},
  doi = {10.1103/PhysRevE.84.061141},
  url = {https://link.aps.org/doi/10.1103/PhysRevE.84.061141}
}

@article{Ezaki_2012,
doi = {10.1088/1742-5468/2012/11/P11002},
url = {https://dx.doi.org/10.1088/1742-5468/2012/11/P11002},
year = {2012},
month = {nov},
publisher = {IOP Publishing and SISSA},
volume = {2012},
number = {11},
pages = {P11002},
author = {Takahiro Ezaki and Katsuhiro Nishinari},
title = {A balance network for the asymmetric simple exclusion process},
journal = {J. Phys. Mech.: Theory Exp.}
}

@article{wang2017dynamics,
  title={Dynamics in multi-lane TASEPs coupled with asymmetric lane-changing rates},
  author={Wang, Yu-Qing and Jia, Bin and Jiang, Rui and Gao, Zi-You and Li, Wan-He and Bao, Ke-Jie and Zheng, Xian-Ze},
  journal={Nonlinear Dyn.},
  volume={88},
  pages={2051--2061},
  year={2017},
  publisher={Springer}
}

@article{wang2018analytical,
  title={Analytical and simulation studies of driven diffusive system with asymmetric heterogeneous interactions},
  author={Wang, Yu-Qing and Wang, Ji-Xin and Li, Wan-He and Zhou, Chao-Fan and Jia, Bin},
  journal={Sci. Rep.},
  volume={8},
  number={1},
  pages={16287},
  year={2018},
  publisher={Nature Publishing Group UK London}
}

@article{Sandow_1994,
doi = {10.1209/0295-5075/26/1/002},
url = {https://dx.doi.org/10.1209/0295-5075/26/1/002},
year = {1994},
month = {apr},
publisher = {},
volume = {26},
number = {1},
pages = {7},
author = {S. Sandow and  G. Schütz},
title = {On Uq[SU(2)]-Symmetric Driven Diffusion},
journal = {Europhys. Lett.}
}

@article{Sarkar_2025,
doi = {10.1088/1742-5468/add514},
url = {https://doi.org/10.1088/1742-5468/add514},
year = {2025},
month = {may},
publisher = {IOP Publishing},
volume = {2025},
number = {5},
pages = {053208},
author = {Sarkar, Mrinal and Gupta, Shamik},
title = {Asymmetric simple exclusion process on a random comb: transport properties in the stationary state},
journal = {Journal of Statistical Mechanics: Theory and Experiment}
}

@article{PhysRevE.69.066128,
  title = {Totally asymmetric exclusion process on chains with a double-chain section in the middle: Computer simulations and a simple theory},
  author = {Brankov, Jordan and Pesheva, Nina and Bunzarova, Nadezhda},
  journal = {Phys. Rev. E},
  volume = {69},
  issue = {6},
  pages = {066128},
  numpages = {13},
  year = {2004},
  month = {Jun},
  publisher = {American Physical Society},
  doi = {10.1103/PhysRevE.69.066128},
  url = {https://link.aps.org/doi/10.1103/PhysRevE.69.066128}
}

@article{Pronina_2005,
doi = {10.1088/1742-5468/2005/07/P07010},
url = {https://doi.org/10.1088/1742-5468/2005/07/P07010},
year = {2005},
month = {jul},
publisher = {},
volume = {2005},
number = {07},
pages = {P07010},
author = {Pronina, Ekaterina and Kolomeisky, Anatoly B},
title = {Theoretical investigation of totally asymmetric exclusion processes on lattices with
junctions},
journal = {Journal of Statistical Mechanics: Theory and Experiment}
}

@article{PhysRevE.77.051108,
  title = {Theoretical investigation of synchronous totally asymmetric exclusion processes on lattices with multiple-input--single-output junctions},
  author = {Wang, Ruili and Liu, Mingzhe and Jiang, Rui},
  journal = {Phys. Rev. E},
  volume = {77},
  issue = {5},
  pages = {051108},
  numpages = {8},
  year = {2008},
  month = {May},
  publisher = {American Physical Society},
  doi = {10.1103/PhysRevE.77.051108},
  url = {https://link.aps.org/doi/10.1103/PhysRevE.77.051108}
}

@article{PhysRevE.80.041128,
  title = {Understanding totally asymmetric simple-exclusion-process transport on networks: Generic analysis via effective rates and explicit vertices},
  author = {Embley, Ben and Parmeggiani, Andrea and Kern, Norbert},
  journal = {Phys. Rev. E},
  volume = {80},
  issue = {4},
  pages = {041128},
  numpages = {22},
  year = {2009},
  month = {Oct},
  publisher = {American Physical Society},
  doi = {10.1103/PhysRevE.80.041128},
  url = {https://link.aps.org/doi/10.1103/PhysRevE.80.041128}
}

@article{Basu_2010,
doi = {10.1088/1742-5468/2010/10/P10014},
url = {https://doi.org/10.1088/1742-5468/2010/10/P10014},
year = {2010},
month = {oct},
publisher = {},
volume = {2010},
number = {10},
pages = {P10014},
author = {Basu, Mahashweta and Mohanty, P K},
title = {Asymmetric simple exclusion process on a Cayley tree},
journal = {Journal of Statistical Mechanics: Theory and Experiment}
}

@article{PhysRevLett.107.068702,
  title = {Totally Asymmetric Simple Exclusion Process on Networks},
  author = {Neri, Izaak and Kern, Norbert and Parmeggiani, Andrea},
  journal = {Phys. Rev. Lett.},
  volume = {107},
  issue = {6},
  pages = {068702},
  numpages = {4},
  year = {2011},
  month = {Aug},
  publisher = {American Physical Society},
  doi = {10.1103/PhysRevLett.107.068702},
  url = {https://link.aps.org/doi/10.1103/PhysRevLett.107.068702}
}

@article{Neri_2013,
doi = {10.1088/1367-2630/15/8/085005},
url = {https://doi.org/10.1088/1367-2630/15/8/085005},
year = {2013},
month = {aug},
publisher = {IOP Publishing},
volume = {15},
number = {8},
pages = {085005},
author = {Neri, Izaak and Kern, Norbert and Parmeggiani, Andrea},
title = {Exclusion processes on networks as models for cytoskeletal transport},
journal = {New Journal of Physics}
}

@article{Mottishaw_2013,
doi = {10.1088/1751-8113/46/40/405003},
url = {https://doi.org/10.1088/1751-8113/46/40/405003},
year = {2013},
month = {sep},
publisher = {IOP Publishing},
volume = {46},
number = {40},
pages = {405003},
author = {Mottishaw, Peter and Waclaw, Bartlomiej and Evans, Martin R},
title = {An exclusion process on a tree with constant aggregate hopping rate},
journal = {Journal of Physics A: Mathematical and Theoretical}
}

@article{PhysRevLett.134.027102,
  title = {Asymmetric Simple Exclusion Process on the Percolation Cluster: Waiting Time Distribution in Side Branches},
  author = {Iyer, Chandrashekar and Barma, Mustansir and Singh, Hunnervir and Dhar, Deepak},
  journal = {Phys. Rev. Lett.},
  volume = {134},
  issue = {2},
  pages = {027102},
  numpages = {7},
  year = {2025},
  month = {Jan},
  publisher = {American Physical Society},
  doi = {10.1103/PhysRevLett.134.027102},
  url = {https://link.aps.org/doi/10.1103/PhysRevLett.134.027102}
}

@article{alexander1992shock,
  title={Shock fluctuations in the two-dimensional asymmetric simple exclusion process},
  author={Alexander, Francis J and Cheng, Zheming and Janowsky, Steven A and Lebowitz, Joel L},
  journal={Journal of statistical physics},
  volume={68},
  number={5},
  pages={761--785},
  year={1992},
  publisher={Springer}
}

@article{Cai_2008,
doi = {10.1088/1742-5468/2008/07/P07016},
url = {https://doi.org/10.1088/1742-5468/2008/07/P07016},
year = {2008},
month = {jul},
publisher = {},
volume = {2008},
number = {07},
pages = {P07016},
author = {Cai, Zhong-Pan and Yuan, Yao-Ming and Jiang, Rui and Hu, Mao-Bin and Wu, Qing-Song and Wu, Yong-Hong},
title = {Asymmetric coupling in multi-channel simple exclusion processes},
journal = {Journal of Statistical Mechanics: Theory and Experiment}
}

@article{SINGH20093113,
title = {Transverse diffusion induced phase transition in asymmetric exclusion process on a surface},
journal = {Physics Letters A},
volume = {373},
number = {35},
pages = {3113-3117},
year = {2009},
issn = {0375-9601},
doi = {https://doi.org/10.1016/j.physleta.2009.06.059},
url = {https://www.sciencedirect.com/science/article/pii/S0375960109007841},
author = {Navinder Singh and Somendra M. Bhattacharjee}
}

@article{SHI20122640,
title = {Strong asymmetric coupling of multilane PASEPs},
journal = {Physics Letters A},
volume = {376},
number = {40},
pages = {2640-2644},
year = {2012},
issn = {0375-9601},
doi = {https://doi.org/10.1016/j.physleta.2012.06.040},
url = {https://www.sciencedirect.com/science/article/pii/S0375960112008365},
author = {Qi-Hong Shi and Rui Jiang and Mao-Bin Hu and Qing-Song Wu}
}

@article{Curatolo_2016,
doi = {10.1088/1751-8113/49/9/095601},
url = {https://doi.org/10.1088/1751-8113/49/9/095601},
year = {2016},
month = {jan},
publisher = {IOP Publishing},
volume = {49},
number = {9},
pages = {095601},
author = {Curatolo, A I and Evans, M R and Kafri, Y and Tailleur, J},
title = {Multilane driven diffusive systems},
journal = {Journal of Physics A: Mathematical and Theoretical}
}

@article{DING20181700,
title = {Analytical and simulation studies of 2D asymmetric simple exclusion process},
journal = {Physica A: Statistical Mechanics and its Applications},
volume = {492},
pages = {1700-1714},
year = {2018},
issn = {0378-4371},
doi = {https://doi.org/10.1016/j.physa.2017.11.091},
url = {https://www.sciencedirect.com/science/article/pii/S0378437117311573},
author = {Zhong-Jun Ding and Shao-Long Yu and Kongjin Zhu and Jian-Xun Ding and Bokui Chen and Qin Shi and Xiao-Shan Lu and Rui Jiang and Bing-Hong Wang}
}

@article{PhysRevResearch.6.L032032,
  title = {Diffusion-facilitated transport of self-driven particles in polycrystalline structures},
  author = {Ezaki, Takahiro and Nishinari, Katsuhiro and Ando, Yasunobu},
  journal = {Phys. Rev. Res.},
  volume = {6},
  issue = {3},
  pages = {L032032},
  numpages = {5},
  year = {2024},
  month = {Aug},
  publisher = {American Physical Society},
  doi = {10.1103/PhysRevResearch.6.L032032},
  url = {https://link.aps.org/doi/10.1103/PhysRevResearch.6.L032032}
}

@article{PhysRevResearch.6.043210,
  title = {Solid-electrolyte-inspired totally asymmetric simple exclusion process with parallel channels and random defects},
  author = {Kihara, Kai and Nishinari, Katsuhiro and Ando, Yasunobu and Ezaki, Takahiro},
  journal = {Phys. Rev. Res.},
  volume = {6},
  issue = {4},
  pages = {043210},
  numpages = {7},
  year = {2024},
  month = {Nov},
  publisher = {American Physical Society},
  doi = {10.1103/PhysRevResearch.6.043210},
  url = {https://link.aps.org/doi/10.1103/PhysRevResearch.6.043210}
}

@article{PhysRevResearch.7.023068,
  title = {Asymmetric simple exclusion process with concerted hopping},
  author = {Ezaki, Takahiro and Kihara, Kai and Nishinari, Katsuhiro and Ando, Yasunobu},
  journal = {Phys. Rev. Res.},
  volume = {7},
  issue = {2},
  pages = {023068},
  numpages = {7},
  year = {2025},
  month = {Apr},
  publisher = {American Physical Society},
  doi = {10.1103/PhysRevResearch.7.023068},
  url = {https://link.aps.org/doi/10.1103/PhysRevResearch.7.023068}
}

@article{mixed,
  title = {A note on mixed hypergeometric series},
  author = {M. A. Khan and A. H. Khan},
  journal = {Acta Math. Vietnam.},
  volume = {14},
  year = {1989},
}

@Article{10.21468/SciPostPhys.17.3.092,
	title={{On perturbation around closed exclusion processes}},
	author={Masataka Watanabe},
	journal={SciPost Phys.},
	volume={17},
	pages={092},
	year={2024},
	publisher={SciPost},
	doi={10.21468/SciPostPhys.17.3.092},
	url={https://scipost.org/10.21468/SciPostPhys.17.3.092},
}

@article{bachman2016inorganic,
  title={Inorganic solid-state electrolytes for lithium batteries: mechanisms and properties governing ion conduction},
  author={Bachman, John Christopher and Muy, Sokseiha and Grimaud, Alexis and Chang, Hao-Hsun and Pour, Nir and Lux, Simon F and Paschos, Odysseas and Maglia, Filippo and Lupart, Saskia and Lamp, Peter and others},
  journal={Chemical reviews},
  volume={116},
  number={1},
  pages={140--162},
  year={2016},
  publisher={ACS Publications}
}

@article{zhao2020designing,
  title={Designing solid-state electrolytes for safe, energy-dense batteries},
  author={Zhao, Qing and Stalin, Sanjuna and Zhao, Chen-Zi and Archer, Lynden A},
  journal={Nature Reviews Materials},
  volume={5},
  number={3},
  pages={229--252},
  year={2020},
  publisher={Nature Publishing Group UK London}
}

@article{famprikis2019fundamentals,
  title={Fundamentals of inorganic solid-state electrolytes for batteries},
  author={Famprikis, Theodosios and Canepa, Pieremanuele and Dawson, James A and Islam, M Saiful and Masquelier, Christian},
  journal={Nature materials},
  volume={18},
  number={12},
  pages={1278--1291},
  year={2019},
  publisher={Nature Publishing Group UK London}
}

@article{10.1063/1.5135319,
    author = {Duan, Chuancheng and Huang, Jake and Sullivan, Neal and O'Hayre, Ryan},
    title = {Proton-conducting oxides for energy conversion and storage},
    journal = {Applied Physics Reviews},
    volume = {7},
    number = {1},
    pages = {011314},
    year = {2020},
    month = {03},
    issn = {1931-9401},
    doi = {10.1063/1.5135319},
    url = {https://doi.org/10.1063/1.5135319},
}

@article{proton2,
   author = "Kreuer, K.D.",
   title = "Proton-Conducting Oxides", 
   journal= "Annual Review of Materials Research",
   year = "2003",
   volume = "33",
   number = "Volume 33, 2003",
   pages = "333-359",
   doi = "https://doi.org/10.1146/annurev.matsci.33.022802.091825",
   publisher = "Annual Reviews",
   issn = "1545-4118",
   type = "Journal Article",
  }

@article{hossain2017review,
  title={A review on proton conducting electrolytes for clean energy and intermediate temperature-solid oxide fuel cells},
  author={Hossain, Shahzad and Abdalla, Abdalla M and Jamain, Siti Noorazean Binti and Zaini, Juliana Hj and Azad, Abul K},
  journal={Renewable and Sustainable Energy Reviews},
  volume={79},
  pages={750--764},
  year={2017},
  publisher={Elsevier}
}

@article{hussain2020review,
  title={Review of solid oxide fuel cell materials: cathode, anode, and electrolyte},
  author={Hussain, Saddam and Yangping, Li},
  journal={Energy Transitions},
  volume={4},
  number={2},
  pages={113--126},
  year={2020},
  publisher={Springer}
}

\end{document}